\documentclass[journal=jacsat,manuscript=article]{achemso}

\usepackage[version=3]{mhchem} 
\usepackage{amsmath, amssymb}
\usepackage{bm}
\usepackage{siunitx}
\usepackage{caption}
\usepackage{subcaption}
\usepackage{booktabs}
\usepackage{longtable}
\usepackage{hyperref}
\usepackage{setspace}
\usepackage{siunitx}

\author{Dun Wang}
\author{Shupeng Xu}
\author{Jia-chen Shi}
\author{Xuyang Li}
\author{Ritesh Agarwal}
\email{riteshag@seas.upenn.edu}
\affiliation[Unknown University]
{Department of Material Science and Engineering, University of Pennsylvania, Philadelphia, USA}

\title[An \textsf{achemso} demo]
  {Geometric Engineering of Flat Bands in a Single-layer Photonic Graphene}

\abbreviations{IR,NMR,UV}

\begin{document}

\begin{tocentry}

Some journals require a graphical entry for the Table of Contents.
This should be laid out ``print ready'' so that the sizing of the
text is correct.

Inside the \texttt{tocentry} environment, the font used is Helvetica
8\,pt, as required by \emph{Journal of the American Chemical
Society}.

The surrounding frame is 9\,cm by 3.5\,cm, which is the maximum
permitted for  \emph{Journal of the American Chemical Society}
graphical table of content entries. The box will not resize if the
content is too big: instead it will overflow the edge of the box.

This box and the associated title will always be printed on a
separate page at the end of the document.

\end{tocentry}

\begin{abstract}
Photonic flat bands offer significant potential for strong light-matter interactions, nonlinear optics, 
and sensing thanks to their localization of light and high density of states. 
However, realizing these flat bands typically requires intricate fabrication, perfect alignment and/or specialized geometries, 
and a general design strategy is missing. 
In this work, we demonstrate a simple yet versatile strategy to engineer radiative flat bands above the light line, 
using only a single-layer honeycomb photonic crystal slab. 
By applying a density wave like geometric perturbation—a spatially periodic displacement of the lattice air holes— 
we couple intrinsic flat band states from below the light cone into the radiative continuum. 
This structural modulation creates a highly anisotropic band structure that exhibits linear, Dirac-like dispersion in one direction 
and nearly flat dispersion in the orthogonal direction, forming an extended van Hove singularity at band extrema. 
Furthermore, by tuning the Fourier components of the modulation, we can manipulate the Dirac mass term to realize band inversion 
and switch between two topologically distinct phases. 
As an application, we demonstrate a Jackiw-Rebbi interface state positioned at the junction of two domains with opposite Dirac mass, 
that also shows flat band dispersion along the interface. 
This density-wave perturbation approach provides a conceptually clear and fabrication friendly platform for programming complex photonic band dispersions, 
opening new avenues for both topological photonics and practical flat-band optoelectronic devices. 
\end{abstract}

\section{Introduction}

Electronic flat bands have provided an extremely fertile playground for correlated phases and topology in condensed matter physics~\cite{cao2018correlated,cao2018unconventional,checkelsky2024flat}. 
Photonic flat bands, as their counterpart, have also garnered much attention as they support ultralow group velocity of light across an extended range of wavevectors~\cite{krauss2008we,leykam2018perspective}. 
This ``slow light" feature has proven useful in enhancing optical nonlinearities~\cite{jiang2025flatband}, optical switching and quantum optics~\cite{krauss2008we}. 
Furthermore, the high density of states and light localization associated with flat bands can facilitate ultrasensitive sensing~\cite{baboux2016bosonic}, enhanced electron radiation~\cite{yang2023photonic} and low threshold lasing~\cite{eyvazi2025flat,do2025room}. 
Significant engineering effort has gone into the realization of flat bands: localized flat band states have been demonstrated in twisted bilayer photonic crystals (PCs)~\cite{dong2021flat,saadi2025tailoring,nguyen2022magic} and moiré-patterned monolayer PCs~\cite{wang2020localization,wang2025one,raun2023gan,mao2021magic}. 
Alternative strategies involve 1D dielectric gratings breaking vertical symmetry~\cite{nguyen2018symmetry,choi2025observation,eyvazi2025flat,do2025room} or special unit cell designs that mimic a zero-index material~\cite{minkov2018zero,tang2021low,cui2025ultracompact}. 
Recently, several novel approaches to realizing flat bands based on artificial Laudau levels~\cite{barsukova2024direct,barczyk2024observation}, bound states in the continuum (BIC)~\cite{qin2025quasi,qi2023steerable, overvig2023zone}, and anapole resonances~\cite{ren2025far} have been put forward and experimentally realized. 
Topological flat bands have been theoretically proposed in coupled optical microcavities with fragile topology~\cite{wang2024design}.
However, many of these approaches either require intricate fabrication like partial etching through the dielectric layer with nm-scale depth control~\cite{choi2025observation,eyvazi2025flat},
or rely on very specific geometries~\cite{minkov2018zero,tang2021low} and perfect alignment of the two layers in the moire case~\cite{dong2021flat,saadi2025tailoring,nguyen2022magic},
which is hard to generalize to different lattice types with different symmetries, thus hindering their practical integration into devices. 

Meanwhile, the design flexibility of single-layer PCs allows tuning of photonic resonances by imposing artificial potentials dictated by the geometric pattern. 
A well known example is the Pancharatam-Berry (PB) phase metasurface~\cite{tymchenko2015gradient,duan2023valley,chen2023compact}, 
where each meta-atom is rotated in a spatially periodic way to effectively generate an additional momentum for circularly polarized light. 
By slightly distorting the underlying air-hole PC lattice or creating an appropriate ``superpotential"~\cite{zhang2025airy}, 
photonic synthetic magnetic fields~\cite{barsukova2024direct,barczyk2024observation} and Airy resonances~\cite{zhang2025airy} have been recently demonstrated. 
The effect of such distortions can be described by Brillouin zone folding if commensurate with the underlying lattice, 
which has been utilized for high-Q resonances~\cite{wang2023brillouin} and enhanced nonlinear frequency conversion~\cite{qin2025enhanced,malek2025giant}. 
Thus, it is intriguing to ask whether adding artificial potentials could facilitate engineering of useful photonic band structures such as flat bands and topological edge states
in simple geometries that are easy to fabricate and can also couple radiatively to the far field.

In this work we demonstrate a simple yet general strategy that realizes simultaneously Dirac and flat dispersion above the light line, 
based on a single photonic analog of graphene-honeycomb PC consisting of air holes etched into a silicon nitride layer. 
This is achieved by modulating the position of the air holes,
where the magnitude of the displacement period is not necessarily commensurate with the underlying lattice.
We call this strategy a density wave-like (DW) perturbation, as it is reminiscent of the charge density wave in condensed matter systems~\cite{gruner1988dynamics}.
By implementing this DW perturbation, we couple the one-dimensional flat band that is present on the lowest band below the light cone 
with the outgoing free space radiation at $\Gamma$ point,
creating a band structure that is linear Dirac-like along $x$ direction and nearly flat along $y$.  
Within the same perturbation scheme the magnitude and sign of the Dirac mass term can also be tuned to realize either of the two topologically distinct phases. 
Finally, we demonstrate a peculiar slow-light Jackiw-Rebbi interface state that exhibits flat dispersion along the domain boundary,
highlighting the versatility of our design for engineering both the band dispersion and topological properties.

\section{Results and discussion}

\begin{figure}
    \centering
    \includegraphics[width=1\textwidth]{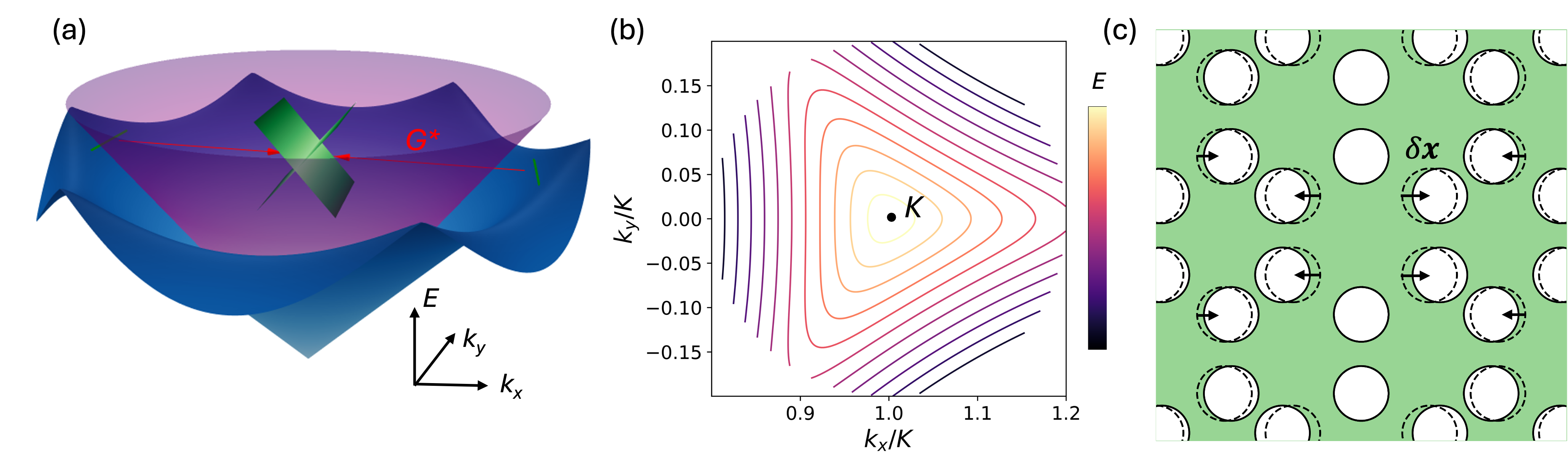}
    \caption{Design principle of our flat band with DW perturbation in a single layer photonic graphene lattice.
    (a) The lowest band of a two-dimensional honeycomb photonic crystal slab is shown in navy blue, which lies below the light cone (the purple surface).
    The solid green lines correspond to the $k_y$ slice where the band curvature along $k_y$ becomes zero at $k_x=\pm G^*$, 
    which is the flat band we are interested in that lies below the light cone. After applying the DW perturbation, the flat band states acquires an additional 
    momentum of $\pm \mathbf{G}^*$, enabling coupling to the far field at $\Gamma$ point. 
    The resulting bands above the light cone are shown in green.
    (b) Isofrequency surface plot of the unperturbed band structure in the vicinity of the $K$ point.
    (c) Schematic of the PCS design, which consists of a silicon nitride layer with circular air holes.
    The solid circles form the unperturbed honeycomb lattice.
    The dashed circles represent the perturbed lattice with a DW perturbation along the $x$ direction.}
    \label{fig_Ch5:fig1} 
\end{figure}

We start with demonstrating the design principle of our PC slab (PCS) with the schematics shown in Figure~\ref{fig_Ch5:fig1}.
The unperturbed band structure of the honeycomb PCS has similar features as the electronic band structure of graphene,
with Dirac points at the $K$ points, around which the dispersion is linear in all directions.
In photonic systems, linear dispersion is also found around the $\Gamma$ point where the dispersion can be described by the effective refractive index in the
deep subwavelength regime as $\omega=\frac{c}{n_{eff}}|\mathbf{k}|$, assuming an isotropic $n_{eff}$.
As a result of the dispersion around $\Gamma$ and $K$ points, the curvature of the energy band along $k_y$ direction
will change sign from positive to negative as one moves from $\Gamma$ to $K$ point along $k_x$.
At a certain point ($G^*$), the curvature will be zero and the band will be flattened along the $k_y$ direction, 
and this is the flat band we seek to exploit.  This idea is illustrated in Figure~\ref{fig_Ch5:fig1}(a), where the lowest band of a honeycomb PCS is shown 
with the 1D flat band highlighted in green, calculated from a coupled plane wave model (see Methods). 
The band structure is for quasi-TE polarization where the electric field vector is mostly in-plane and 
hereafter we will only focus on this polarization. A more detailed view of the band structure in the vicinity of $K$ point is provided as an isofrequency surface plot
in Figure~\ref{fig_Ch5:fig1}(b), where the flat band is found at $k_x\equiv G^*\approx 0.92 G_K$, $G_K=\frac{4\pi}{3a}$ being the wavevector for the $K$ point,
and the inversion of curvature can be clearly seen its two sides.
The flat part has an extension $\Delta k_y$ of more than $0.2G_K$ in $y$ direction and the ``flatness'' only varies slightly for small deviation from $G^*$, 
implying robustness against experimental imperfections.
The exact value of $G^*$ in PCs has to be determined numerically, and can deviate significantly from 
the nearest-neighbour tight-binding model prediction of $G^*=\frac{3}{4}G_K$ for graphene\cite{mcchesney2010extended}, as the photonic coupling can occur over a much longer range than in electronic systems.

As long as the time-reversal symmetry is preserved, the photonic band structure is always symmetric 
in $\mathbf{k}$ and $-\mathbf{k}$\cite{joannopoulos2008molding}, thus the flat band appears at both $k_x=+G^*$ and $-G^*$ (see Figure~\ref{fig_Ch5:fig1}).
The flat band lies below the light cone ($\omega=c|\mathbf{k}|$, purple surface in Fig.~\ref{fig_Ch5:fig1}(a)) and is not accessible in the far field. To couple it 
with the radiative continuum, we apply the DW modulation which shifts the position of the air hole at site $\mathbf{R_i}$ by an amount 
$\delta\mathbf{x_i}=\mathbf{\tau_1}\sin(\mathbf{G}^* \cdot \mathbf{R_i})$, as shown in Figure~\ref{fig_Ch5:fig1}(c). Writing the original dielectric potential of a honeycomb PCS as 
$V_0(\mathbf{r})=\sum_i v(\mathbf{r}-\mathbf{R_i})$ where $v(\mathbf{r})$ is the potential of a single lattice site and the sum is over all lattice vectors,
the change in the potential landscape due to the DW modulation is represented by 
\begin{equation}
  V_0(\mathbf{r}) \rightarrow V(\mathbf{r})=\sum_i v(\mathbf{r}-\mathbf{R_i}-\mathbf{\tau_1}\sin(\mathbf{G}^* \cdot \mathbf{R_i}))
  \label{eq1}
\end{equation}
For small $\tau_1$, the perturbation from the DW modulation can be shown (see Supporting Information) to have the following form:
\begin{equation}
  \delta V(\mathbf{r}) = - |\boldsymbol{\tau_1}\cdot\mathbf{G}^*| \tilde{v}(\mathbf{G}^*) \sin(\mathbf{G}^* \cdot \mathbf{r} ) + \mathcal{O}(\tau_1^2)
  \label{eq2}
\end{equation}
where $\tilde{v}$ is the Fourier transform of $v(\mathbf{r})$. From Eq.~\ref{eq2} it is apparent that the DW modulation creates a 
potential with dominant Fourier component $\pm\mathbf{G}^*$, which enables coupling between the flat band at $k_x=\pm G^*$ and free space plane waves at $\Gamma$.

We first verify our design using numerical simulations (see Methods). The geometry consists of a silicon nitride (SiN) layer of thickness $h=120$ nm,
with etched circular air holes forming a honeycomb lattice. 
The lattice constant is $a=265$ nm and
the diameter of the air holes is $d=100$ nm.
Beneath the silicon nitride layer is a SiO$_2$/Si substrate with SiO$_2$ thickness of 800 nm; 
this thickness is chosen to optimize the contrast of the reflectance spectrum with regard to the Fabry-Perot resonances in the substrate (See Supporting Information). 
The eigenmodes of the unperturbed system are first calculated using COMSOL Multiphysics, 
where we identified that the flat band is located around $\mathbf{G}^* = 0.925\mathbf{G}_K$ away from the $\Gamma$ point, for the parameters specified above.
It should be stressed that in general the $G^*$ modulation does not have to be commensurate with the underlying lattice
and our design principle does not rely on a periodic supercell approximation.
We then add the DW perturbation by displacing the air holes along $x$ direction with a magnitude of $\tau_1 = 20$ nm and a wavevector of $\mathbf{G}^*$,
and calculate the reflectance spectrum in the far field.
For this task, we use the finite-difference time-domain (FDTD) solver Tidy3D\cite{tidy3d}.
Here, and later in experiments, we focus on reflection from linearly polarized light along $y$ as the flat band states in concern are dominantly $y$ polarized.

\begin{figure}
\centering
\includegraphics[width=1\textwidth]{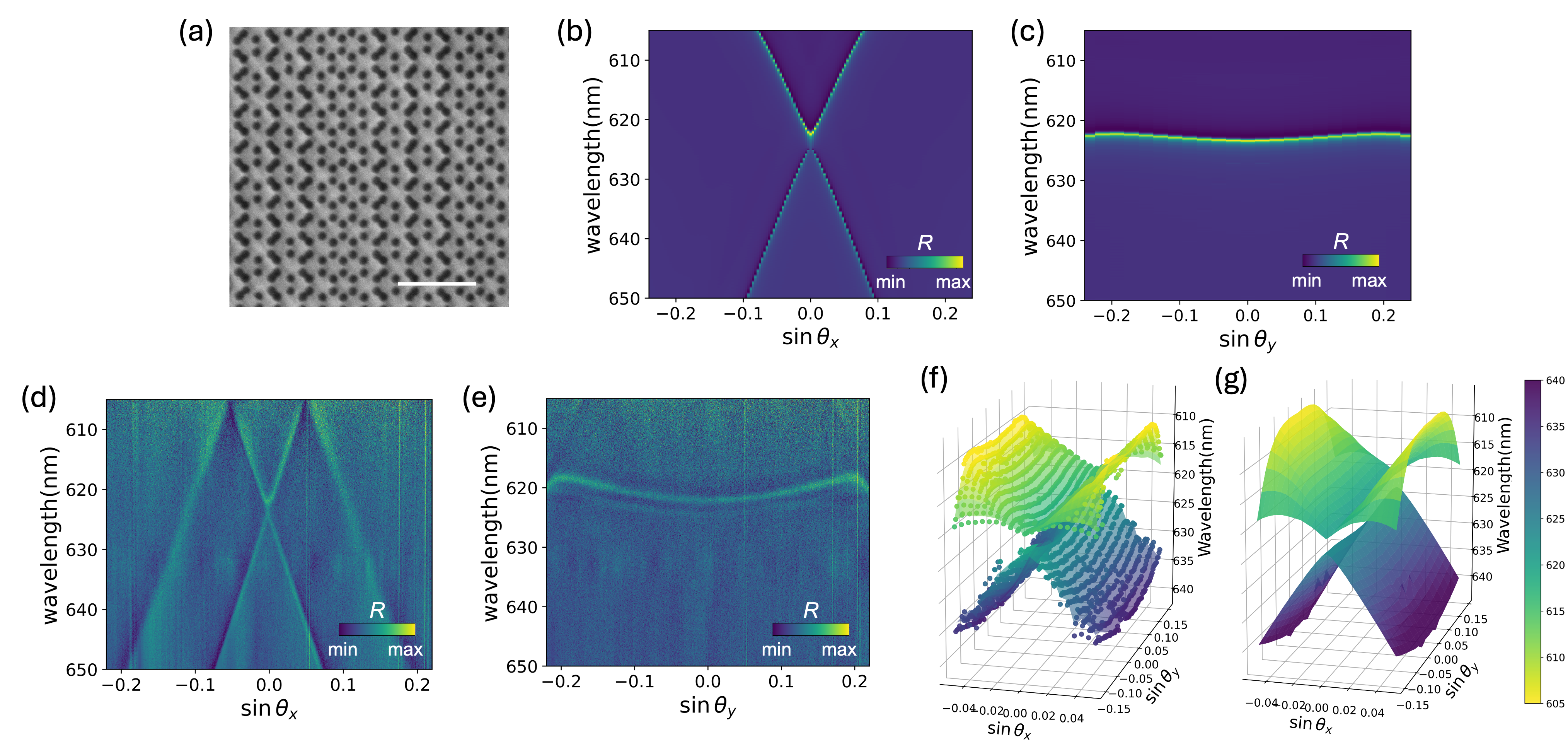}
\caption{Band structure of the photonic graphene lattice with a DW modulation.
    (a) SEM image of the fabricated PCS sample with the DW modulation applied. The scale bar is 1 $\mu$m.
    (b)(c) Numerical simulation of the angle resolved reflection spectrum of the PCS, along (b) $x$ and (c) $y$ directions, respectively.
    The dispersion in the $x$ direction is linear and Dirac-like with a small gap at $\Gamma$ point, 
    while the $y$ dispersion is nearly flat with only one bright band. 
    (d)(e) Experimental angle resolved reflection spectrum of the PCS, along (d) $x$ and (e) $y$ directions, respectively.
    (f)(g) 3D tomographic reconstruction of the band structure from (f)experimental and (g)simulated reflection spectrum. 
    In (f), the dots follow the peak positions extracted from experimental data and the surface plot is a polynomial fit of the bands. 
    Both figures are false colored by the $z$ height, which corresponds to the wavelength.}
  \label{fig_Ch5:fig2} 
\end{figure}

The numerically calculated reflectance spectra (Figure~\ref{fig_Ch5:fig2}b, c) shows that
along the $x$ direction, the band has Dirac-like dispersion, while along the $y$ direction, the band is nearly flat with a bandwidth of about 1 nm extending
between $\sin \theta_y = \pm0.2$, corresponding to a wide angular range of $\Delta \theta_y=23^\circ$. 
Such a band structure implies an effective mass that is near-zero along $x$ but approaching infinite along $y$,
exhibiting extreme anisotropy.
The Dirac dispersion along $x$ is slightly gapped at $\Gamma$, with the upper branch being bright and the lower dark branch supporting a bound state in the continuum (BIC).
This behavior is characteristic of the Friedrich-Wintgen BIC mechanism\cite{friedrich1985interfering,hsu2016bound}, 
where two resonances couple to the same radiation channel 
and interference between the radiative states leads to the via-the-continuum coupling.

Consider two resonances $|\psi_1\rangle$ and $|\psi_2\rangle$ with eigenfrequencies $\omega_1$ and $\omega_2$, 
and decay rates $\gamma_1$ and $\gamma_2$, respectively.
The effective Hamiltonian for the system can be written as
\begin{equation}
H_{\text{eff}} = \begin{pmatrix}
\omega_1  & \kappa \\
\kappa & \omega_2 
\end{pmatrix} - i \begin{pmatrix}\gamma_1 & \sqrt{\gamma_1\gamma_2}\cos\phi \\
\sqrt{\gamma_1\gamma_2}\cos\phi & \gamma_2 \end{pmatrix},
\label{bic}
\end{equation}
where $\kappa$ represents the near-field coupling between the two resonances,
and the magnitude of the via-the-continuum coupling is given by $\sqrt{\gamma_1\gamma_2}\cos\phi$, with $\phi$ being the difference in polarization angle.
In our case, $|\psi_1\rangle, |\psi_2\rangle$ stands for a pair of flat band states at $(k_x=\pm G^*, k_y)$ below the light cone, 
with $\omega_1 = \omega_2 \equiv \omega_0$. Their radiative decay rates $\gamma_1 = \gamma_2 \equiv \gamma$ comes from 
the DW modulation which scatters them to the far field at $\Gamma$ point and are proportional to $\tau_1$. 
The eigenstates of Hamiltonian Eq.~\ref{bic} are given by $|\psi_{\pm}\rangle = \frac{1}{\sqrt{2}}(|\psi_1\rangle \pm |\psi_2\rangle)$, 
with the eigenvalues $\omega_{\pm} = \omega_0 \pm \kappa - i(\gamma \pm \gamma\cos\phi)$.
When the two states are co-polarized so that $\phi=0$, the eigenvalue for $|\psi_-\rangle$ is purely real, corresponding to the BIC state with infinite lifetime;
for small nonzero $\phi$, the radiative loss are not cancelled completely and $|\psi_-\rangle$ would become a quasi-BIC state. 
For a wide range of $k_y$, $|\psi_1\rangle$ and $|\psi_2\rangle$ are very close to being linearly co-polarized along $y$ (see Supporting Information),
resulting in the lower band with negligible radiation to the far field across the entire angular range.
Whether $|\psi_-\rangle$ or $|\psi_+\rangle$ is higher in energy depends on the sign of $\kappa$.
The gap in the real part of $\omega_{\pm}$, as is evident in Figure~\ref{fig_Ch5:fig2}(b), implies a non-zero $\kappa$ that 
couples states at $G^*$ and $-G^*$ and thus must come from a $2G^*$ Fourier component in the perturbation potential in Eq.~\ref{eq2}. 
Such a term can come from expanding $V(\mathbf{r})$ to second order in $\tau_1$, 
giving a term proportional to $\tau_1^2\sin^2(\mathbf{G}^* \cdot \mathbf{r})$ which contains a $2G^*$ component via trigonometric relations.

Experimentally, we deposited the SiN and SiO$_2$ layers with the designed thickness 
and patterned the PCS sample using the standard dry etching process (see Methods).
The SEM image of the sample is shown in Figure~\ref{fig_Ch5:fig2}(a). 
Measured angle-resolved reflection spectrum (Figure~\ref{fig_Ch5:fig2}d, e) 
shows the desired linear dispersion along $x$ and nearly flat dispersion along $y$, matching well with the numerical results. 
The gap is not resolved in the experiments due to a slight increase in the linewidth owing to fabrication imperfections. 
The extra linearly dispersive features in Figure~\ref{fig_Ch5:fig2}(d) are from the guided modes in the finite-thick SiO$_2$ layer, 
which are not seen in simulations where the substrate is assumed to be half-infinite.
In order to have a complete picture of the band structure, we measured the reflection spectrum in a range of $\theta_x$ and $\theta_y$,
and extracted the 3D band structure from both experiment and simulation data (Figure~\ref{fig_Ch5:fig2}f and g, respectively).
For better visualization, the experimental spectrum at each $\theta_x$ are fitted with a sixth order polynomial 
$\omega(k_y)=\omega_0+ c_2 k_y^2 + c_4 k_y^4 + c_6 k_y^6$, where $k_y=2\pi/\lambda\sin\theta_y$. 
The sixth order term is needed because these flat bands appear near the curvature inversion point where $c_2$ is close to zero, 
therefore $c_4k_y^4$ is the lowest order in dispersion and $c_6k_y^6$ captures the deviation from flatness at larger $k_y$.

\begin{figure}
\centering
\includegraphics[width=1\textwidth]{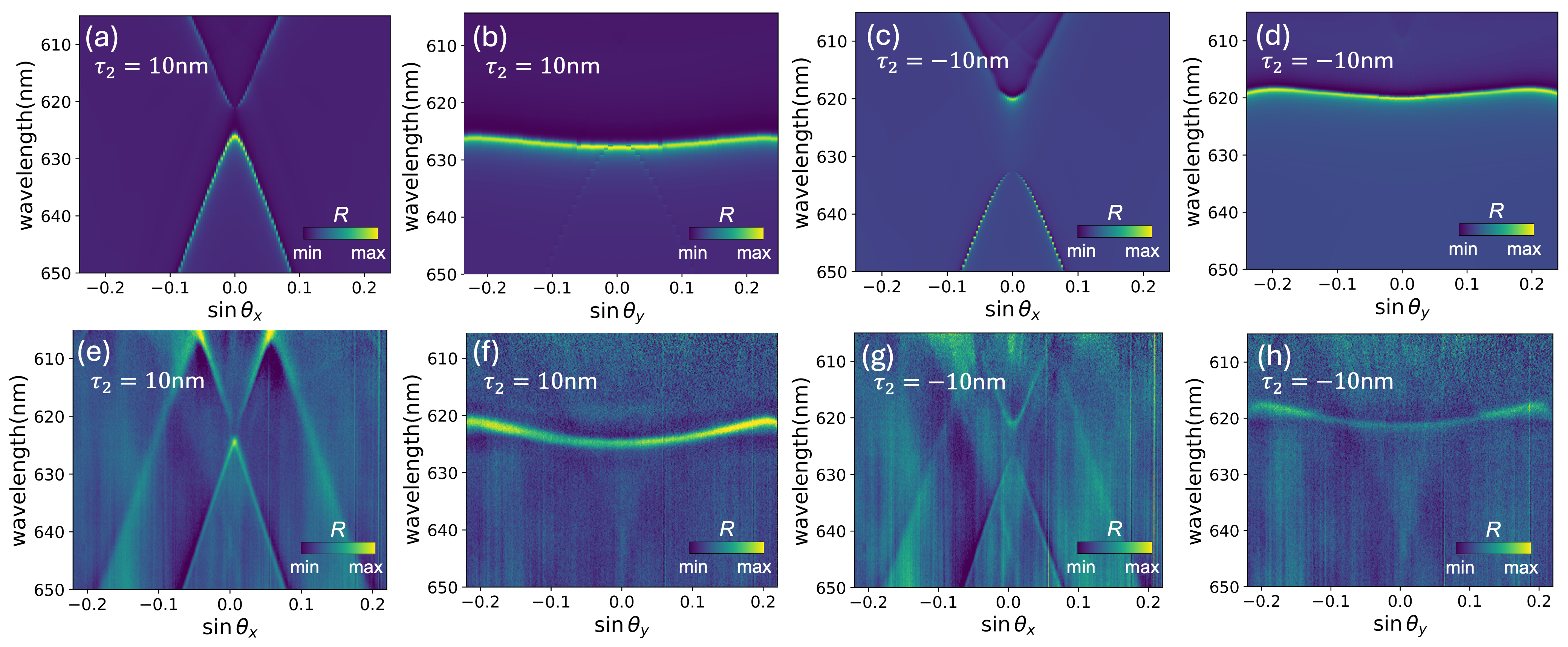}
\caption{Topological bandgap engineering from the DW modulation.
    (a-d) Simulated and (e-h) experimental angle resolved reflection spectrum of the PCS with $\tau_1=20$ nm and $\tau_2 = \pm 10$ nm, 
    along $x$ and $y$ directions, as indicated in each panel.
    }
  \label{fig_Ch5:fig3} 
\end{figure}

The $\kappa$ term in Eq.~\ref{bic} can be easily engineered in our perturbation scheme by adding an additional modulation with a wavevector of $2G^*$. 
To achieve this, we introduce a second parameter $\mathbf{\tau_2}$ into the positional shift of the air holes in Eq.~\ref{eq1}:
\begin{equation}
  V_0(\mathbf{r}) \rightarrow V(\mathbf{r})=\sum_i v(\mathbf{r}-\mathbf{R_i}-\mathbf{\tau_1}\sin(\mathbf{G}^* \cdot \mathbf{R_i})-\mathbf{\tau_2}\sin(2\mathbf{G}^* \cdot \mathbf{R_i}))
  \label{eq3}
\end{equation}
Not only the magnitude but also the sign of $\kappa$ can be changed by changing the sign of $\tau_2$, 
leading to a reversal of the bright and dark states as the upper or lower branch. 
This is demonstrated numerically and experimentally in Figure~\ref{fig_Ch5:fig3}, where the band structures with $\tau_1=20$ nm and $\tau_2=\pm 10$ nm are shown.
Along the $x$ direction, 
when $\tau_2 = 10$ nm, the BIC state is located in the upper branch, while when $\tau_2 = -10$ nm, the BIC state is supported by the lower branch.
In both cases, the reflection spectrum shows that the bands along $y$ direction remain nearly flat, 
and only one band is visible, indicating that the BIC state is still maintained along the entire $k_y$ slice. 
With the gap opened, the flat band consititutes a so-called extened van Hove singularity at $\Gamma$, 
where the band curvature vanishes in one direction in addition to the usual saddle point van Hove singularity \cite{mcchesney2010extended,gofron1994observation}. 
Extended van Hove singularities have stronger divergence in density of states than conventional van Hove singularities and plays an important role in increasing the critial temperature of superconductivity in various materials\cite{radtke1994relation, yuan2019magic};
in photonics, this stronger divergence can enhance band-gap confinement\cite{ibanescu2006enhanced} and increase the hybridization strength for cavity-polariton formation\cite{gianardi2025formation}. 
The gap size is larger for $\tau_2 = -10$ nm than $\tau_2 = 10$ nm, due to the aforementioned second order contribution of the $\tau_1$ term to $\kappa$
which favors gap opening with BIC in the lower branch.

The easily customizable Dirac mass term in our PCS design provides a natural platform for engineering
topological interface states.
Unlike conventional waveguide or cavity confinement, topological interface states arise solely from the
topological distinction between two bulk phases, making it robust against disorder and
fabrication imperfections that would otherwise scatter guided modes\cite{ozawa2019topological}.
As an application of our design, we demonstrate a Jackiw-Rebbi (JR) interface state with slow light along the domain boundary.
The JR state was first introduced in the context of quantum field theory\cite{jackiw1976solitons},
where it was shown that the 1D Dirac equation with a spatially varying mass term $m(x)$ that changes sign at $x=0$,
\begin{equation}
(-i v_F \sigma_x \partial_x + m(x) \sigma_z) \psi(x) = E \psi(x),
\end{equation}
supports a zero-energy solution localized at the domain boundary.
On both sides of the domain, the wavefunction decays exponentially as $\psi(x) \sim e^{-\int_0^x m(x') dx'}$.
The existence of the zero mode has a topological origin,
where it is guaranteed by the change in the sign of the mass term across the interface,
corresponding to a change in the topological invariant defined by the Zak phase.
This mode cannot be removed by any perturbation as long as the bulk gap on both sides of the domain remains open.
The JR mechanism underlies many topological phenomena such as the Su-Schriffer-Heeger (SSH) edge state 
and has been observed in various condensed matter systems such as polyacetylene, topological insulators, and others~\cite{hasan2010colloquium}.
In photonic systems, it has been realized in various platforms including 1D SSH-like structures and 
2D topological photonic crystals~\cite{randerson2025topological,choi2025topological,gorlach2019photonic,gupta2024realization,liu2020z2},
among which a particularly simple realization is by creating an interface between two one-dimensional gratings with a band inversion~\cite{randerson2025topological,lee2022topological}.

\begin{figure}
\centering
\includegraphics[width=1\textwidth]{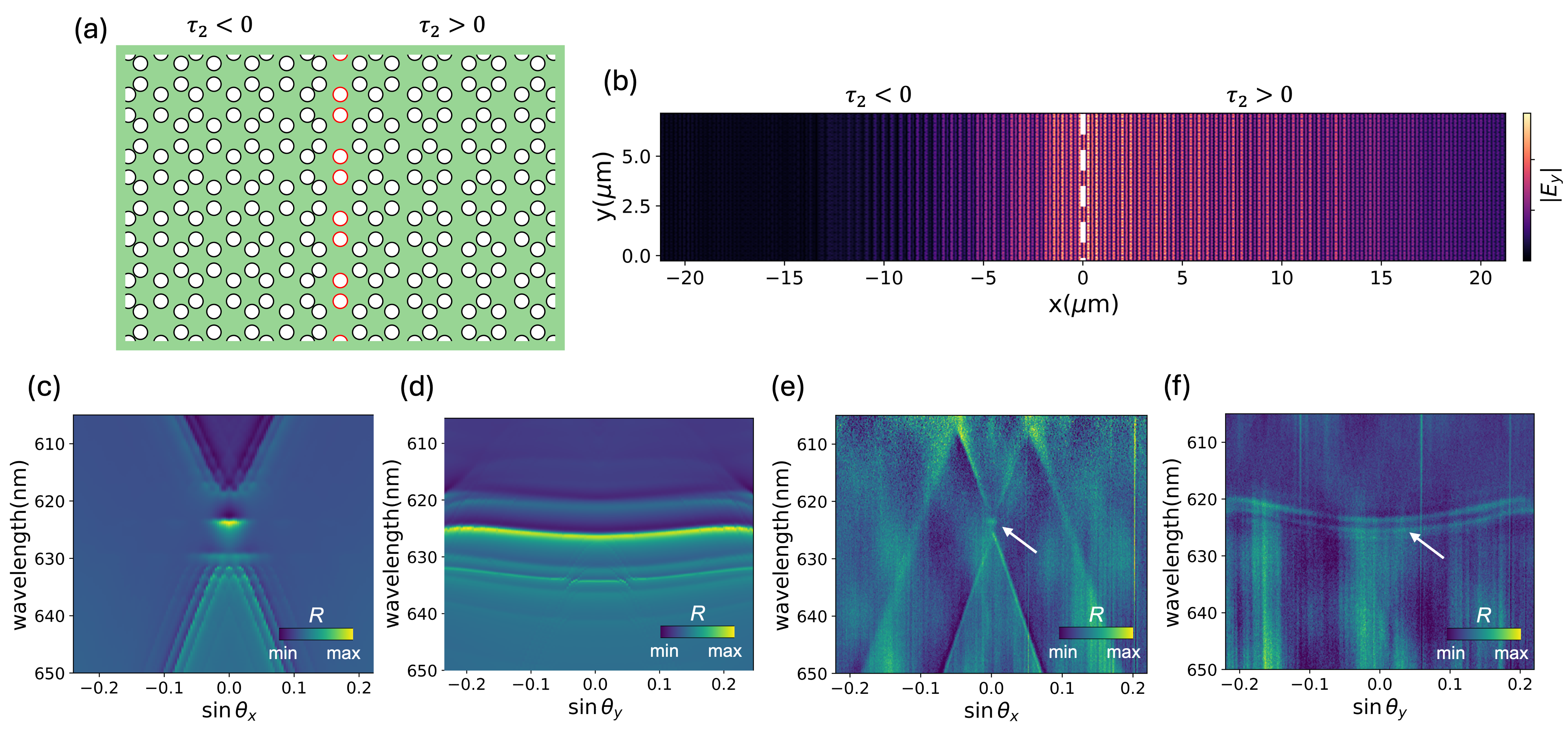}
\caption{Schematic and characterization of the flat band Jackiw-Rebbi state.
    (a) Real space geometry of the JR interface state. The domain boundary is marked in red.
        $\tau_2$ changes sign from -10 nm to 10 nm across the boundary.
    (b) Simulated electric field distribution for the JR interface state.
        The white broken line marks the domain boundary at $x=0$.
    (c,d) Simulated and (e,f) experimental angle resolved reflection spectrum measured at the JR interface, along (c,e)$x$ and (d,f)$y$ directions, respectively.
    The JR state can be identified as the bright reflectance feature in the middle of the gap, pointed by the white arrow.    
    }
  \label{fig_Ch5:fig4} 
\end{figure}

In our system, the JR interface is created by changing the sign of $\tau_2$ across the boundary at $x=0$, as shown in Figure~\ref{fig_Ch5:fig4}(a).
Along the $y$ direction, the translational symmetry is intact and the nearly flat band dispersion is expected to be the same as the bulk.
The simulated electric field profile of the JR state (Figure~\ref{fig_Ch5:fig4}b) shows that
the field is localized at the boundary and decays exponentially into the bulk,
and has a longer decay length on the right side of the domain boundary where $\tau_2$ is positive and the gap is smaller (cf. Figure~\ref{fig_Ch5:fig3}a, c).
The calculated and experimentally measured reflection spectrum along $x$ and $y$ directions (Figure~\ref{fig_Ch5:fig4}c-f) 
shows the JR state clearly realized in the middle of the gap. 
In the $k_y$ resolved spectrum (Figure~\ref{fig_Ch5:fig4}d, f), three nearly flat bands can be identified: the middle one corresponds to the JR state and
the upper and lower parts are from the bulk bands on two sides of the interface. 
The experimental bandgap is smaller than the simulated gap owing to fabrication imperfections slightly blurring the perturbation potential, 
but the JR state can be clearly resolved. The spatial field localization along $x$ and the slow light feature along $y$ lead to 
increased local density of states and power density, which could be beneficial for nonlinear optical processes including four wave mixing and optical parametric amplification.

In conclusion, we have introduced a density-wave-like geometric perturbation of a honeycomb PCS that provides a powerful 
and fabrication-friendly handle for engineering both the band dispersion and far-field properties of photonic resonances. 
We demonstrated numerically and experimentally a variety of exotic features including 
flat and Dirac dispersion within a highly anisotropic band structure, band inversion with BICs and extended van Hove singularities, and 
topological Jackiw-Rebbi interface states on a flat band. The perturbation parameters $\tau_1$ and $\tau_2$ are connected straightforwardly
to the outcoupling with far field radiation and the gap opening between the $\pm G^*$ states, respectively,
offering a flexible handle for optimizing the design for different applications. 
The bands show flat dispersion along $y$ for $\sin\theta_x$ ranging between $\pm 0.06$, which can be useful as a platform to study 1D physics 
such as the bosonic Luttinger liquid\cite{haller2010pinning,bouchoule2025platforms}.

Taking a broader perspective, the DW modulation approach we introduced here establishes a general design principle, 
where a geometric perturbation of the lattice acts as a synthetic potential in reciprocal space.
With this method, one can fold bands, open gaps, and redistribute oscillator strength between dark and bright modes through a small set of tunable parameters.
More complicated synthetic potentials can be programmed by adding more Fourier components in the perturbation, 
allowing for the exploration of many interesting physics such as moire photonics~\cite{mao2021magic}, 
structured light~\cite{zhang2020tunable} and topological defect states~\cite{song2025artificial} in a single PCS geometry.
Extending to the incommensurate and non-perturbative regime would lead into the field of quasicrystals, 
where facinating new phenomena including high topological charge lasing~\cite{arjas2024high,che2021polarization} 
and high dimensional topological invariants~\cite{tsesses2025four} have been recently observed.
Our findings open new avenues for engineering photonic band structures
and could enable novel applications in light-matter interaction and topological photonics.

\subsection{Methods}

\subsubsection{Sample Fabrication.}
The PCS samples were fabricated on SiO\textsubscript{2}/Si substrates, 
where the thickness of the silicon dioxide layer is 285 nm.
We first deposited a silicon dioxide layer using Oxford PlasmaLab 100 by plasma-enhanced chemical vapor deposition (PECVD) to make the total thickness of silicon dioxide to be 800 nm for better optical contrast in the reflectance measurement.
A 120 nm silicon nitride layer was then deposited on top of the substrate using the same method.
A 2 minute oxygen plasma treatment was performed to improve the adhesion of the resist,
then the sample was spin-coated with ZEON ZEP520A-7 positive electron beam resist at 2000 r.p.m. for 60 seconds,
which yields a uniform resist layer of 300 nm.
The sample was then baked at $\SI{180}{\degreeCelsius}$ for 3 minutes to remove the solvent.
The pattern was defined by electron beam lithography (EBL) using Raith EBPG5200 system at 100 kV acceleration voltage and 10 nA beam current and an aperture size of 300 $\mu$m.
The electron beam spot size was around 10 nm, and the beam step size 5 nm.
The shape detection mode was used to improve the pattern fidelity,
and the pattern was written at a dosage of 600 $\mu$C/cm$^2$,
which was corrected by proximity effect correction. 
After exposure, the sample was developed in cold o-Xylene for 60 seconds followed by rinsing in isopropanol for 30 seconds and drying with nitrogen gas. 
The pattern was then transferred to the silicon nitride layer by reactive ion etching (RIE) using Oxford PlasmaLab 100 with CHF\textsubscript{3} and O\textsubscript{2} gases.
Finally, the sample was cleaned by sonication in hot N-Methyl-2-pyrrolidone (NMP), acetone and isopropanol for 5 minutes each, followed by drying with nitrogen gas.
Lastly, the sample was cleaned by O\textsubscript{2} plasma for 5 minutes to remove any residual hardened resist and contaminants.

\subsubsection{Numerical Simulations.}
The eigenmodes of the unperturbed PCS was calculated using the Electromagnetic Wave, Frequency Domain interface in COMSOL Multiphysics. 
A hexagonal primitive unit cell was simulated with Floquet periodicity boundary conditions, and thick PMLs were added in the $\pm z$ directions.
The polarization of the eigenmodes was evaluated by integrating the cell-periodic part of the electric field over the unit cell and dividing by the unit cell area.
The refractive index of SiN and SiO\textsubscript{2} were set to be 2.02 and 1.457, respectively.
For calculating the reflectance spectrum of the modulated system, the FDTD solver Tidy3D was used as it is more suitable for large-area simulations.
The system was excited by a plane wave source linearly polarized along $y$ from above and the reflected light was collected by a flux monitor.
The FDTD simulation region contained 60 honeycomb unit cells along $x$ direction in order to make the system periodic in the presence of the DW modulation,
and the Bloch boundary condition was applied along $x$ and $y$ direction. 
This was to avoid additional periodicity from finite size of the simulation area, and we note again that the DW modulation does not have to be commensurate with the lattice.
In the simulations, the SiO$_2$ substrate was assumed to be half-infinite and the bottom silicon was neglected to save computational resources, 
which does not affect the results qualitatively as the SiO$_2$ layer is optically thick enough.
The runtime was set to 840 fs to ensure that the incident pulse had fully decayed.
The reflectance spectrum was then calculated by sweeping the incident angle.
For reflection from the JR state, we set the source with a finite extension of 26.5 $\mu$m centered at the domain boundary for large mode overlap with the interface state.
A field monitor was placed at the center of the slab to record the electric field profile of the JR state.

\subsubsection{Theoretical Model.}
The lowest band structure of the single-layer honeycomb PCS can be described by a simple coupled plane wave model. 
We take as basis the plane waves originating from the first few reciprocal lattice vectors, with their dispersion given by $\omega_i(\mathbf{k}) = \frac{c}{n_{eff}}|\mathbf{k}+\mathbf{G_i}|$,
where $\mathbf{G_i}$ is a reciprocal lattice vector~\cite{joannopoulos2008molding}. 
To capture the essential features of the lowest band in the irreducible Brillouin zone,
only the three sets of plane waves originating from $\mathbf{G_0}=(0,0), \mathbf{G_1}=(\frac{2\pi}{a}, \frac{2\pi}{\sqrt3 a}), \mathbf{G_2}=(\frac{2\pi}{a}, -\frac{2\pi}{\sqrt3 a})$
are needed. Assuming a constant coupling strength $t$ between them, the Hamiltonian can be written as
\begin{equation}
H = \begin{pmatrix}
\omega_0(\mathbf{k}) & t & t \\
t & \omega_1(\mathbf{k}) & t \\
t & t & \omega_2(\mathbf{k})
\end{pmatrix}
\label{model}
\end{equation}
Diagonalization of Eq.~\ref{model} gives the band structure shown in Figure~\ref{fig_Ch5:fig1}. 
The flat band wavevector $G^*$ depends on $t$ and we chose $t=0.035\frac{c}{n_{eff}}|G_K|$ to match with our numerical and experimental results.

\subsubsection{Measurement Setup.}

Measurements of the angle-resolved reflection spectrum were performed with a home-built Fourier space spectroscopy setup. 
Broadband white light was focused onto the sample via a 10X, 0.25 NA objective (Nikon), which also collected the back-reflected light. 
The objective's back focal plane was projected directly onto the slit of the spectrometer (Acton SpectraPro 500i). 
The spectrometer’s EMCCD (Andor, Oxford Instruments) simultaneously recorded the spectral distribution and the spatial position, 
which maps to the angular distribution of the light entering the slit.

\begin{acknowledgement}

This work was supported by the
Office of Naval Research under Grant No. N00014-22-
1-2378 and by the National Science Foundation through
the DMREF program (Grant No. NSF-2323468).
The authors thank Utkarsh Khandelwal for his help with the SEM. 
The authors also thank Flexcompute Inc. for providing access to the Tidy3D software used for all FDTD
simulations in this work.

\end{acknowledgement}

\begin{suppinfo}

Detailed derivation of Equation ~\ref{eq2}; Discussion of effects from the silicon substrate;
Polarization data of the flat band states; Additional experimental and numerical data.

\end{suppinfo}

\bibliography{chapter5_bib.bib}

\end{document}